\documentclass{aa}

\usepackage{psfig}
\begin{document}

   \thesaurus{03  
              (11.01.2;   
               11.19.1;   
               11.09.1;   
               11.11.1;   
               11.19.6)}  

   \title{Kinematics and morphology of the Narrow-Line Region 
          in the Seyfert galaxy NGC\,1386
 \thanks{Based on observations collected at the European Southern 
        Observatory, La Silla, Chile}}

   \subtitle{}

   \author{J.~Rossa
         \inst{1,2}
          \and
         M.~Dietrich
          \inst{1,3}     
          \and   
         S.~J.~Wagner
          \inst{1}}
   \offprints{jrossa@astro.ruhr-uni-bochum.de}
   
   \institute{Landessternwarte Heidelberg--K\"onigstuhl, K\"onigstuhl 12, 
             D--69117 Heidelberg, Germany
         \and 
             Astronomisches Institut der Ruhr-Universit\"at Bochum,
              D--44780 Bochum, Germany\thanks{\emph{Present address}} 
         \and
             Department of Astronomy, University of Florida, 211 Bryant Space 
             Center, Gainesville, FL 32611-2055, USA$^{\star\star}$}

   \date{Received 29 March 2000 / Accepted 15 August 2000}

   \maketitle

   \begin{abstract}

We present a high spatial and spectral resolution 2-D echelle spectrogram of 
the Narrow--Line Region in the Seyfert\,2 galaxy \object{NGC\,1386}. This 
Seyfert galaxy was observed with CASPEC in the wavelength range 
5270--7725\,{\AA} which covers the H$\alpha$ and the [\ion{N}{ii}] lines. With 
the use of spatially high resolved images taken with the WFPC2 aboard the 
Hubble Space Telescope we could identify individual components of the 
Narrow--Line Region in our spectra. A Gaussian decomposition of the spectra 
revealed 9 distinct emission--line complexes. The brightest component is 
blue-shifted by $-120\pm10\,\mathrm{km\,s^{-1}}$ with respect to the systemic 
velocity and shows an offset of $\approx-1\farcs6$ relative to the nucleus of 
the galaxy. The true nucleus of NGC\,1386 has a much lower apparent H$\alpha$ 
luminosity than this component. The nucleus is probably highly absorbed. 
Although the majority of the Narrow--Line Region components follows a regular 
velocity field, we find evidence for a separate kinematic component. The 
Narrow--Line Region is aligned anti--parallel to the radio--jet which 
propagates from the center of NGC\,1386 to the south.

   \end{abstract}

%

\section{Introduction}

The emission line regions of active galactic nuclei (AGN) can be 
divided into the Broad--Line Region (BLR) and the Narrow--Line Region (NLR). 
The study of velocity fields of the narrow emission--line region in nearby 
AGN offers an exciting perspective to study details of the circumnuclear 
kinematics. Besides the primary component, which is given by the rotation of 
the galaxy, additional velocity components are likely to be present. 
Especially, in AGN velocity components, such as inflow or outflow motions, 
may be superimposed onto the regular rotational velocity field 
(e.g.,\,Osterbrock \cite{Osterbrock}). The morphology of the ionized gas 
component often coincides with radio structures, seen in VLA maps 
(e.g.,\,Ulvestad \& Wilson \cite{UW}; Pogge \cite{RWP}). Such extended radio 
structures (e.g.,\,radio jets) are often aligned with the optical emission 
(cf.~Capetti et al. \cite{Capetti3}; Gallimore et al. \cite{GBD}). 

First hints for a complex structure of the emission line region of AGN were 
mentioned by Walker (quoted by Burbidge et al. \cite{BBP}) in NGC\,4151 und 
NGC\,1068. In the late 60s and later on particularly for NGC\,1068 strong 
indication for a structured emission line region und indiviual substructures 
were found (Walker \cite{W68}; Anderson \cite{A70}; Glaspey et al. \cite{
G76a},\,b; Alloin et al. \cite{APBS}). Recent investigations of nearby Seyfert 
galaxies with high spatial resolution, as it is achieved with the Hubble 
Space Telescope (HST), showed that it is feasible to resolve individual NLR 
components in the nearest Seyfert galaxies (e.g.,\,\object{NGC\,1068}, 
Macchetto et al. \cite{Macchetto}; \object{NGC\,4151}, Winge et al. 
\cite{Wi97}; \object{Mkn\,3}, \object{Mkn\,78}, \object{Mkn\,348}, Capetti 
et al. \cite{Ca95}, \cite{Capetti2}) on scales as small as 0\farcs1. In the 
case of NGC\,1068 the morphology of the NLR can be characterized as a clumpy 
structure of several dozen individual emission line knots.

\subsection{NGC\,1386}

\begin{figure}[h]
\psfig{figure=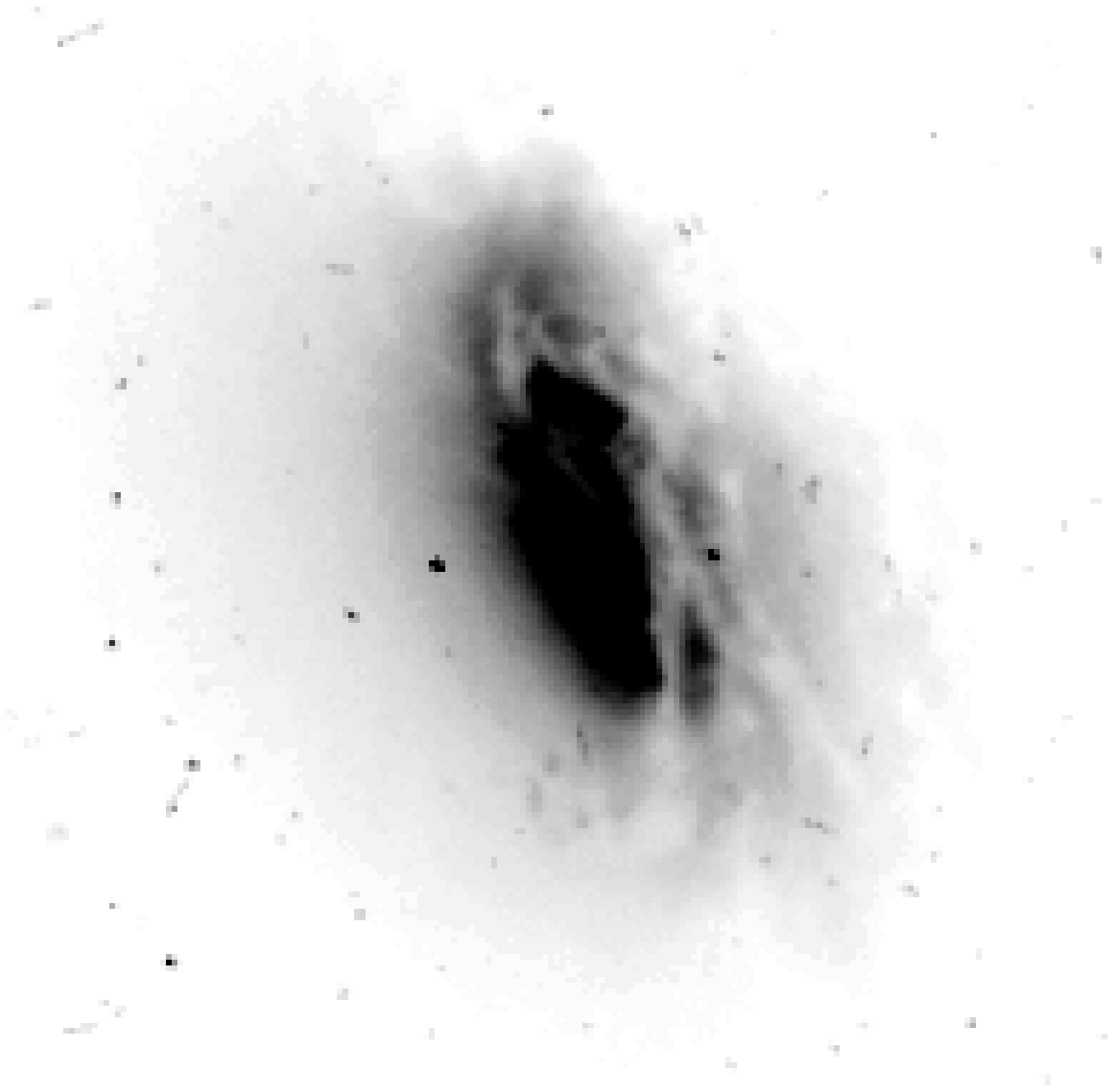,width=8.8cm,clip=t}
\caption[]{The Seyfert\,2 galaxy NGC\,1386. This is the central region 
of the HST archive image obtained with the WFPC\,2, and was taken with 
the F606W filter. North is up, East is to the left. The dust lanes, which run 
across NGC\,1386, are easily recognizable.}
\label{F1}
\end{figure}

The galaxy \object{NGC\,1386} (Fig.~\ref{F1}) has been classified as a Seyfert 
type 2 galaxy by Phillips \& Frogel (\cite{PF}). The spectra show 
characteristic strong emission lines of H$\alpha$, H$\beta$, [\ion{N}{ii}], 
and [\ion{O}{iii}] with expected line ratios and line widths typical for a 
Seyfert nucleus.

\begin{table*}[t]
\caption[]{Basic data on NGC\,1386}
\label{T1}
\begin{flushleft}
\begin{minipage}{20cm}\small 
\begin{tabular}{llllllllll} 
\noalign{\smallskip}
\hline
Galaxy & R.~A.\,(J\,2000)\footnote{All data have been taken or have been 
calculated from the RC\,3 (de Vaucouleurs et al. \cite{Vau91}), except where 
indicated} & Dec.\,(J\,2000) & Hubble type\footnote{taken from Weaver et al. 
(\cite{WWB})} & AGN type & $D$ & $v_{\rm{\ion{H}{i}}}$\footnote{taken from 
Huchtmeier \& Richter (\cite{HR})} & $a\times b$ & $i$ & 
$m_{\rm{R}}$\footnote{taken from the NASA Extragalactic Database (NED)} \\ 
\hline
\noalign{\smallskip}
NGC\,1386 & ${\rm{03^h36^m46\fs4}}$ & $-35\degr59\arcmin58\farcs1$ & Sa & 
Sy\,2 & 12.2\,Mpc & $918\,{\rm km\,s^{-1}}$ & $3\farcm5\times1\farcm3$ & 
$71\degr$ & 11.24 \\
\noalign{\smallskip}
\hline
\end{tabular}
\end{minipage}
\end{flushleft}
\end{table*}

There is some debate on the morphological galaxy type, whether it is Sa or 
S0, due to the relatively high inclination of $i \approx 71\degr$. Tully 
(\cite{Tully}) lists it as S0 while generally a Sa classification is assumed 
(e.g.,\,Sandage \& Tammann \cite{ST}; Weaver et al. \cite{WWB}). The 
assumption from the Weaver et al. study was based on the line ratios found 
for the outer regions of NGC\,1386 which were characteristic of \ion{H}{ii} 
regions. In the recent literature one can also find classifications with SB0+ 
(Tsvetanov \& Petrosian \cite{TP}), although there are no clear indications 
for the existence of a bar.

\vspace{0.5cm}
\vspace{0.5cm}

Since attention was first paid on this galaxy, there was some uncertainty 
whether NGC\,1386 belongs to the Fornax galaxy cluster or not. Huchtmeier \& 
Richter (\cite{HR}) list a heliocentric corrected systemic velocity of 
$918\pm34\,\mathrm{km\,s^{-1}}$, derived from \ion{H}{i} measurements which 
is in agreement with the value of $890\pm20\,\mathrm{km\,s^{-1}}$ obtained by 
Weaver et al. (\cite{WWB}), as derived from the rotation curve. 

We assume a Hubble parameter, that will be used throughout this paper, of
$\mathrm{H_0}=75\,\mathrm{km\,s^{-1}\,Mpc^{-1}}$. Assuming no effect from the 
Fornax Cluster the radial velocity, obtained by Huchtmeier \& Richter 
(\cite{HR}), translates to a distance to NGC\,1386 of 12.2\,Mpc. Basic data 
on NGC\,1386 are listed in Table~\ref{T1}. 

The distance to the Fornax Cluster has been determined by McMillan et al. 
(\cite{McMillan}), using the Planetary Nebula Luminosity Function (PNLF). 
From their study of three member galaxies of the Fornax Cluster they derived 
a distance of $16.9\pm1.1\,\mathrm{Mpc}$. This value is coincident with the 
value that Tully (\cite{Tully}) listed in his Nearby Galaxies Catalog. More 
recent measurements using Cepheid based distances yielded $18.6\pm0.6$\,Mpc 
(Madore et al. \cite{Ma99}). They also derived a mean velocity of the Fornax 
Cluster of $v=1441\,\mathrm{km\,s^{-1}}$ with $\sigma=\pm342\,\mathrm{km\,
s^{-1}}$. We therefore assume that NGC\,1386 is located in the foreground of 
the Fornax Cluster. 

Being one of the closest Seyfert nuclei, NGC\,1386 is an ideal candidate for 
high spectral and spatial resolution investigations of the NLR, and we 
therefore chose NGC\,1386 for a detailed investigation of the NLR kinematics 
and morphology. One of the techniques applied to this object class besides 
narrowband imaging (e.g.,\,Evans et al. \cite{Evans}; Lynds et al. 
\cite{Lynds}; Pogge \& De\,Robertis \cite{Pogge}) is the speckle masking 
method (e.g.,\,Mauder et al. \cite{Mauder}), although there are alternative 
techniques that have already been applied to the nearest Seyfert galaxies 
successfully (e.g.,\,Wagner \& Appenzeller \cite{SWIA}; Dietrich \& Wagner 
\cite{DW}). However, the first attempt to resolve individual NLR components 
within NGC\,1386 was made using speckle masking observations (Mauder et al. 
\cite{Mauder}). From their study it was inferred that the morphology of 
the gas distribution is cone--like.

Weaver et al. (\cite{WWB}) studied the NLR in NGC\,1386 with medium spectral 
resolution. They reported on the velocity field as a combination of a 
normally rotating component and a component undergoing high velocity infall 
or outflow, as derived from the shape and width of the line profiles. The 
line width amounts to $540\pm40\,\rm{km\,s^{-1}}$ at the nucleus with a 
double peak line profile. These measurements are in agreement with our better 
resolved line profiles (due to higher spectral and spatial resolution) which 
yield a FWHM of $470\pm20\rm{km\,s^{-1}}$. Furthermore, they reported that 
the optical nucleus is displaced from the kinematical center of the rotation 
curve by $1\arcsec-1\farcs5$ which corresponds to 60--90\,pc.


\section{Observations and data reduction}

\subsection{Echelle spectroscopy}

The Seyfert galaxy NGC\,1386 was observed at ESO La Silla, Chile (Nov. 16, 
1992) with the Cassegrain Echelle Spectrograph (CASPEC) attached to the ESO 
3.6m telescope. The achieved spectral resolution with the short camera is 
$\mathrm{R}=\frac{\lambda}{\Delta\lambda}\approx 18000$ which corresponds to 
$\Delta v = 17\,\mathrm{km\,s^{-1}}$. With the use of the red cross-disperser 
the echelle spectrum covers the range 5270--7725\,{\AA}. The slit covered an 
area of $8\arcsec \times1\arcsec$ and was oriented at p.a. $16\degr$. The CCD 
chip, a TEK\,512M--12 with $512\times512$ pixels of 27$\mu$m\,$\rm{pix}^{-1}$
offers a spatial scale of $\rm{0\farcs65\,pix^{-1}}$. The NGC\,1386 spectrum
had an integration time of 5400\,sec.

The data reduction was performed with the IRAF\footnote{IRAF is distributed 
by the National Optical Astronomy Observatories, which is operated by the 
Association of Universities for Research in Astronomy, Inc. (AURA) under 
cooperative agreement with the National Science Foundation.} and 
MIDAS\footnote{Munich Image Data Analysis System, distributed by ESO} 
software packages. The spectrum was bias corrected creating a master bias 
which was subtracted from the spectrum. The tracing was determined from 
standard star exposures. In this process a Chebychev polynomial of 
$3^{\mathrm{rd}}$ order was fitted along the dispersion axis of each 
aperture. The master trace was defined by the spectrum with the smallest 
rms deviations. The echelle spectrum was then background corrected making 
use of the interorder sections of the 2D echelle spectrogram. 

The flatfield correction was done in two different steps. Due to spatial 
offsets and different slit lengths, we had to create several groups of 
flatfields. The internal scatter of the spatial offsets was less than 1\,pix 
within a flatfield group. The deviation between the individual flatfield 
groups had offsets of the order of $\Delta y \approx$ 2\,pix. For each group 
a normalized flatfield was computed which were used to correct for overall 
sensitivity variations of the corresponding science frames. Due to the 
flexure of the spectrograph the echelle object spectra had small offsets with 
respect to the corresponding flatfield exposure. Hence, we were forced to 
restrict the 2D flatfield correction to the inner 3\arcsec while the 1\farcs5 
wide stripes above and below this region were not corrected in this step. The 
normalized flatfield was set to unity for the region of these stripes. The 
criterium for the selection of these scans based on an intensity of less than 
$\approx$50\,\% of the maximal intensity on the position of the individual 
scan. In the second step the final extracted 1D spectra of the inner 6\arcsec 
of the individual spatial scans were flatfield corrected with a 1D flatfield 
generated. It was created by normalizing the individual scans of the used 
flatfield for the 2D correction with the average 1D flatfield of it. Finally 
we are able to make use of the 2D echelle spectrograms covering the NLR gas 
of NGC\,1386 out to a radius of $\approx3\arcsec$ from the nucleus. 

Subsequently the data were wavelength calibrated by using a ThAr lamp 
reference spectrum. The unit step was set to 0.1464\,{\AA}. The scatter of 
our wavelength calibration had an rms of 0.014\,{\AA}. The spectral resolution 
of our spectra was obtained from measurements of several night--sky lines in 
our spectra and was found to be 0.3\,{\AA}. A final flux calibration was 
applied using the standard HR\,3454 ($\eta$\,Hya) which was reduced in the 
same way as the 2-D spectrum of NGC\,1386. The extracted spectrum was finally 
merged and a two-dimensional image was created from the 11 pixel scans. To 
correct for the effect of slit illumination, we divided our 2-D galaxy 
echelle--spectrogram by an identically processed normalized sky--flatfield 
exposure. Finally a cosmics correction was applied to the spectrum using the 
MIDAS {\sl{modify/gcurs}} routine.

\begin{figure}[h]
\psfig{figure=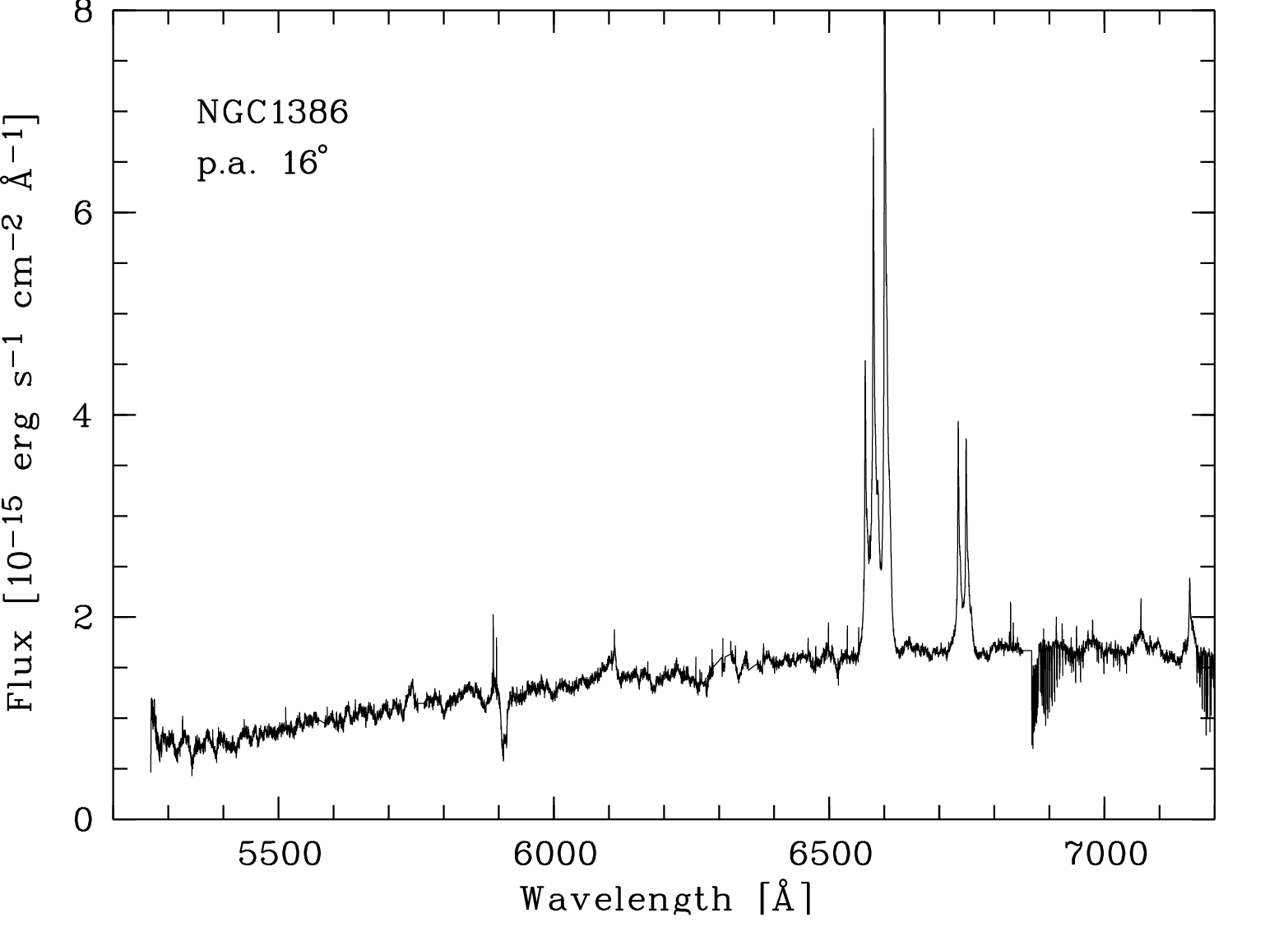,width=8.8cm,clip=t}
\caption[]{The averaged, central $2\farcs6$ of the NGC\,1386 2\,D echelle
spectrogram are shown. The flux in units of $10^{-15}\,
\mathrm{erg\,s^{-1}\,cm^{-2}}$\,{\AA}$^{-1}$ is plotted versus the 
wavelength in the range of 5270--7200\,{\AA}. The interstellar Na\,D 
absorption line at $\lambda \approx 5900$\,{\AA} is clearly visible.}
\label{F2}
\end{figure}

In Fig.~\ref{F2} we present the spectrum of NGC\,1386 from the central 
$2\farcs6$. The typical strong emission lines of Seyfert galaxies, 
e.g.,\,H$\alpha$, [\ion{N}{ii}], [\ion{S}{ii}] are visible. Above 6900\,{\AA} 
the spectrum is dominated by strong atmospheric absorption bands (b-band, 
A-band).

\vspace{0.5cm}
\vspace{0.5cm}

\subsection{Imaging}

We retrieved high resolution images of NGC\,1386 from the Space Telescope 
Science Institute (STScI) data archive (from the survey by Malkan et 
al.~\cite{MGT}, Prop.--No.\,5479). This image had been obtained with the 
WFPC\,2 aboard the Hubble Space Telescope (HST). It was taken with the F606W 
filter ($\lambda_{mean}$=5843.3\,{\AA}, FWHM=1578.7\,{\AA}). Further details 
are given in Biretta et al.~(\cite{Biretta}). 

Additionally, further WFPC\,2 images of NGC\,1386 became available from a 
survey of Seyfert galaxies carried out by Wilson et al. (Prop.--No.\,6419). 
These images had been taken in two narrowband and two broadband filters. In 
our study we made use of the F547M and F658N filter images. The narrowband 
(F658N) image covers the H$\alpha$+[\ion{N}{ii}] emission within the filter 
FWHM of 28.5\,{\AA}. The WFPC\,2 images of NGC\,1386 were used to identify the 
individual components of the NLR in our 2-D spectrum which will be described 
in detail in Sect. 4.3.


\section{Analysis}

In addition to the pure emission line fluxes the calibrated spectrum also 
contains the radiation from the underlying stellar population (continuum). To 
remove this effect in a simple approach we calculated a linear continuum fit 
to our spectrum across the emission lines of interest (e.g.,\,H$\alpha$, 
[\ion{N}{ii}]). This linear interpolation which based on the mean flux in 
continuum windows ($\approx$10\,{\AA} width) at either side of the 
H$\alpha$+[\ion{N}{ii}] doublet emission line complex 
($\lambda_{blue}^c$=6538\,{\AA}, $\lambda_{red}^c$=6675\,{\AA}) was 
determined for each spatial scan and subtracted.

The 2-D echelle spectrogram was transformed from wavelength space into 
velocity space as described by Dietrich \& Wagner (\cite{DW}). We used the 
heliocentric velocity $v\mathrm{_{sys}=918\pm34\,km\,s^{-1}}$ of NGC\,1386, 
as given by Huchtmeier \& Richter (\cite{HR}). We checked this value by 
measuring the centroid of the redshifted interstellar Na\,I line and applying 
the appropriate heliocentric correction. This results in $v\mathrm{_{\rm sys} 
= 924\pm51\,km\,s^{-1}}$, in good agreement with Huchtmeier \& Richter 
(\cite{HR}). 

Since our spectrum covers the wavelength range 5270--7725{\AA}, 
we have several strong emission lines which are useful for the kinematical 
study of the line emitting gas. We use H$\alpha$ and the [\ion{N}{ii}] lines 
for our kinematical investigation. The less intense 
[\ion{S}{ii}]\,$\lambda\lambda$\,6717,6731 doublet, 
[\ion{O}{i}]\,$\lambda$\,6300 and the 
[\ion{Fe}{vii}]\,$\lambda\lambda$\,5721,6087 lines will be used for the 
investigation of the spatial variations of the physical conditions in the 
individual NLR emission--line clumps.


\section{Results}

\subsection{Kinematics}

\begin{figure}[h]
\psfig{figure=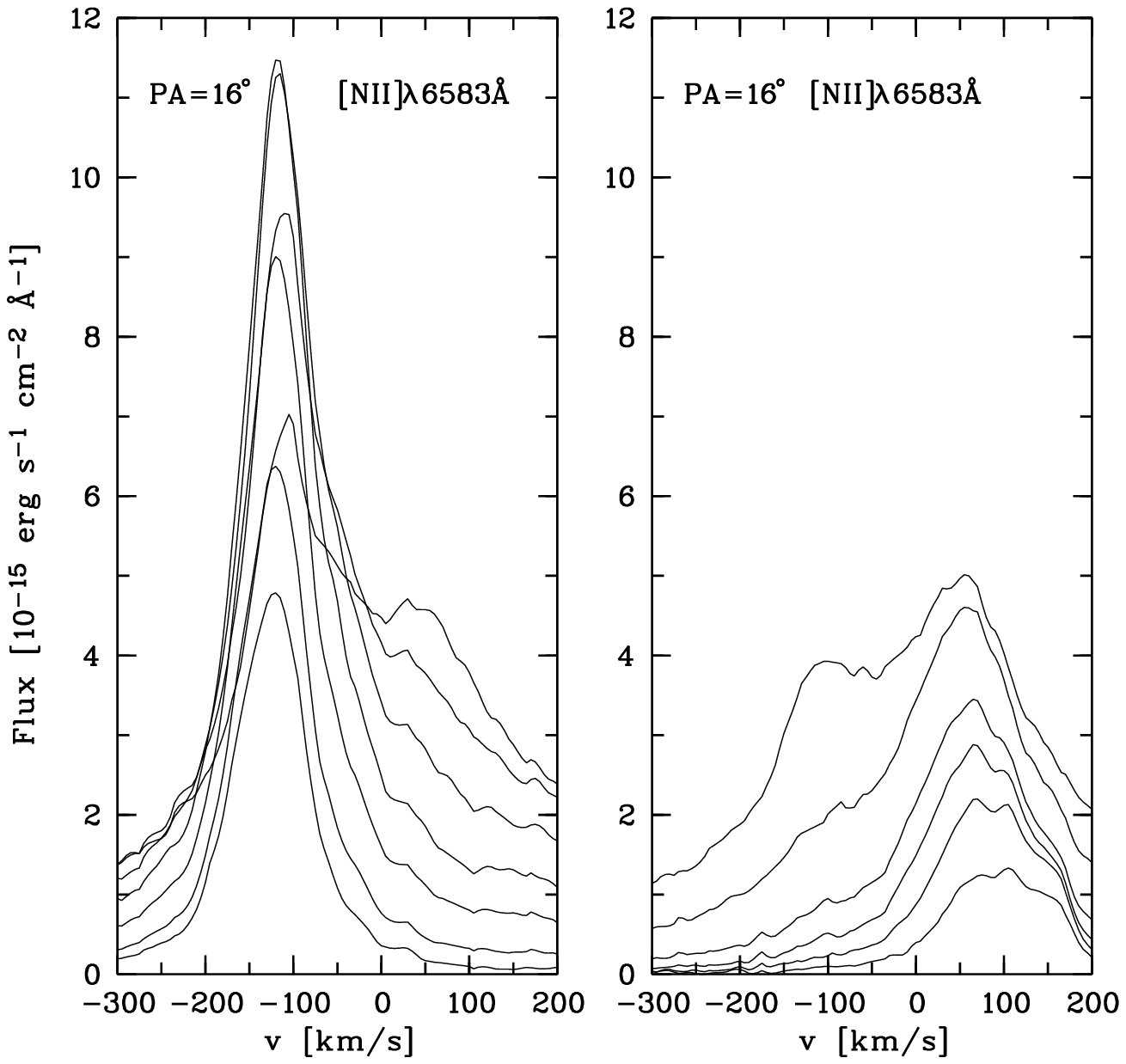,width=10.8cm,clip=t}
\caption[]{The spatial profiles of the [\ion{N}{ii}]\,$\lambda$\,6583\,{\AA} 
line are shown. The velocity relative to the systemic velocity is plotted 
versus the flux in units of $10^{-15}\,\mathrm{erg\,s^{-1}\,cm^{-2}}$\,{\AA}
$^{-1}$. The left panel shows 6 southern scans plus the center scan, each 
separated by $0\farcs5$ from the nucleus, and the right panel shows 6 
northern scans, also separated from one another by $0\farcs5$.}
\label{F3}
\end{figure}

By transforming the spectrum into the velocity space, the $l-v$ map shown in 
Fig.~\ref{F5}\,c has been obtained. At a distance of 12.2\,Mpc, the area 
covered by the slit of the spectrograph projected onto the plane of the 
galaxy is $480\,\mathrm{pc} \times 60\,\mathrm{pc}$. The smallest details 
visible in the $l-v$--map are estimated to be of the order of 0\farcs3 which 
corresponds to a linear size at the distance of NGC\,1386 of only 
$\sim 18\,\mathrm{pc}$. With this high spatial resolution in our spectra it 
is possible to obtain a detailed picture of the kinematics of the NLR in 
NGC\,1386, based on a 2-D echelle spectrum with spatial information as a 
function of velocity. In Fig.~\ref{F3} the spatial scans of the velocity space 
transformed spectrum in the [\ion{N}{ii}]\,$\lambda$6583 line are shown. In 
order to make a quantitative approach, a decomposition using elliptical 
Gaussians was applied to the spectrum. This yielded the position and velocity 
of the individual components in the $l-v$--map as well as the line widths for 
each individual cloud in the $l-v$ space. The detailed description of the 
process of decomposing a 2-D echelle spectrogram is given by Dietrich \& 
Wagner (\cite{DW}) for the Seyfert galaxy NGC\,1068.

\vspace{0.5cm}
\vspace{0.5cm}

The decomposition of the NGC\,1386 spectra yielded 9 individual NLR clouds 
which show mean velocity differences up to 400\,${\rm{km\,s^{-1}}}$ in 
[\ion{N}{ii}]\,$\lambda$6583. The individual NLR components, marked by 
letters, and derived by the Gaussian decomposition (see Fig.~\ref{F5}\,c) are 
located at different velocities and spatial locations relative to the nucleus. 
The line widths of these clouds vary from about $90-235\,{\rm{km\,s^{-1}}}$ 
(clouds A, D, F-I) and are accompanied by clouds with considerably higher 
values of FWHM = $270-470\,{\rm{km\,s^{-1}}}$ (clouds B, C, and E). The 
largest and most prominent region in this $l-v$--map is component A which is 
blueshifted with respect to the systemic velocity by 
$-120\pm10\,\mathrm{km\,s^{-1}}$ and shows an offset of $-1\farcs6$ relative 
to the nucleus. 

\begin{table*}[t]
\caption[]{Positions and velocities of the NLR components in H$\alpha$ and 
[\ion{N}{ii}]\,$\lambda$6583}
\label{T2}
\begin{flushleft}
\begin{tabular}{lcccccccc} 
\noalign{\smallskip}
\hline
&\multicolumn{4}{c}{H$\alpha\,\lambda$6563}
&\multicolumn{4}{c}{[\ion{N}{ii}]\,$\lambda$6583}\\
Component & $v\,[\mathrm{km\,s^{-1}}]$ & FWHM\,$[\mathrm{km\,s^{-1}}]$ & 
Pos.\,[\arcsec]\footnotemark[3] & FWHM\,[\arcsec] & $v\,[\mathrm{km\,s^{-1}}]$ & FWHM\,$[
\mathrm{km\,s^{-1}}]$ & Pos.\,[\arcsec] & FWHM\,[\arcsec]\\ 
at P.A. $16\degr$ & & & & & & & & \\
\hline
\noalign{\smallskip}
A & $-125\pm10$ & 99 & $-1.60$ & 2.35 & $-120\pm10$ & 89 & $-1.60$ & 2.35\\ 
B & $-200\pm10$ & 400 & $-0.40$ & 2.24 & $-235\pm10$ & 471 & $-0.35$ & 2.24\\ 
C & $+200\pm10$ & 353 & $-0.35$ & 2.47 & $+200\pm10$ & 377 & $-0.33$ & 2.47\\ 
D & $+75\pm10$ & 130 & $+1.95$ & 2.59 & $+75\pm10$ & 153 & $+1.95$ & 2.59\\  
E & $-145\pm10$ & 271 & $-0.25$ & 2.59 & $-155\pm10$ & 271 & $-0.35$ & 2.59\\ 
F & $+570\pm10$ & 283 & $-0.30$ & 2.35 & $+580\pm10$ & 235 & $-0.30$ & 2.35\\
G & $-50\pm10$ & 118 & $-0.85$ & 2.47 & $-35\pm10$ & 141 & $-0.85$ & 2.47\\
H & $-95\pm10$ & 95 & $-0.20$ & 2.12 & $-80\pm10$ & 118 & $-0.20$ & 2.35\\
I & $+35\pm10$ & 106 & $+0.60$ & 1.88 & $+53\pm10$ & 113 & $+0.50$ & 1.88\\
\noalign{\smallskip}
\hline
\end{tabular}
\end{flushleft}
\footnotetext*{$^3$~all positional accuracies are of the order of 0\farcs1.}
\end{table*}

We have plotted the velocities versus the position in Fig.~\ref{F5}\,c 
with the derived uncertainties, which show the various velocity components 
for the [\ion{N}{ii}]\,$\lambda$6583 line. The analysis for H$\alpha$ 
provided nearly identical results.    

The identified components with their respective velocities (relative to 
the systemic velocity) and their positions in our $l-v$--map are listed 
both for H$\alpha$ and [\ion{N}{ii}]\,$\lambda$6583 in Table~\ref{T2}. The 
uncertainties arise from the positional accuracies of our $l-v$--map. The 
velocity uncertainties are of the order of $\Delta\,v\approx10\,{\rm{km\,s^{
-1}}}$ and in spatial direction of $\Delta\,l\approx0\farcs1$.
  
\subsection{Morphology of the NLR}

Several emission lines of different ionization levels are covered by the
wavelength range of our 2-D echelle spectrogram. Therefore, it is possible
to search for differences in the morphology of the extended line emission
as a function of the ionization potential. With this method one can roughly 
discriminate NLR components from luminous \ion{H}{ii} regions since  
\ion{H}{ii} regions are not supposed to be visible in the high ionization 
lines as [\ion{Fe}{vii}] for instance. The mechanism for exciting the gas to 
temperatures of that degree, where highly ionized Fe--lines become apparent, 
e.g.,\,[\ion{Fe}{vii}], can not be achieved with common stellar continuum 
emission as for instance in a massive starburst, since starbursts can not 
supply such high temperatures. Therefore, the most likely process is 
photoionization of the gas by a nonstellar power law continuum provided by 
the central continuum source or by continuum emission of fast radiative 
shocks (e.g.,\,Dopita \& Sutherland \cite{DS1}). The usual discrimination 
between \ion{H}{ii} regions, AGN--like, and starburst galaxy regions is 
described in the diagnostic diagrams by plotting the intensity ratios of 
different emission lines 
(e.g.,\,$\frac{\mathrm{[\ion{O}{iii}]}}{\mathrm{H}\beta}$ vs. 
$\frac{\mathrm{[\ion{N}{ii}]}}{\mathrm{H}\alpha}$, cf.~Baldwin et al. \cite{
BPT1}; Veilleux \& Osterbrock \cite{VeOs}; Allen et al. \cite{ADT}).  

Luminous \ion{H}{ii} regions in the vicinity of an Seyfert nucleus are 
frequently found. These regions are indicative of massive starbursts. 
Sometimes the \ion{H}{ii} regions appear in a ring structure around 
the nucleus, which is seen for instance in \object{NGC\,7469} (De\,Robertis 
\& Pogge \cite{DP}) and NGC\,1068 (Balick \& Heckman \cite{BH}). In NGC\,1386 
there are 44 \ion{H}{ii} regions visible which were detected by Tsvetanov \& 
Petrosian (\cite{TP}) in their survey of \ion{H}{ii} regions in Seyfert 
galaxies. These reside in larger radii to the nucleus, outside our studied 
central region. 

The brightest component which we refer to as cloud A, was initially suspected 
to be an ultraluminous \ion{H}{ii} region (Rossa \cite{Rossa96}), based on 
the morphology of the 2-D echelle line profiles of different excitation 
levels. The reason that NGC\,1386 is a low luminosity Seyfert galaxy and a 
detection of an apparent ultraluminous \ion{H}{ii} region may lead to the 
conclusion that NGC\,1386 is a starburst galaxy. However, the detection of the 
high ionization lines, which are characteristic for AGN, rules out that 
possibility. Furthermore, the line ratio [\ion{N}{ii}]\,$\lambda$6583/H$\alpha 
\simeq$2 rules out the possibility even that component A is a \ion{H}{ii} 
region. It turns out that component A is located in the classical region for 
power law excitation in the Seyfert regime, according to diagnostic diagrams 
(e.g.,\,Baldwin et al. \cite{BPT1}). 

The comparison of the morphology of the NLR in the Seyfert\,2 galaxy NGC\,1386 
with each of the low ionization lines (e.g., H$\alpha$, [\ion{N}{ii}], 
[\ion{S}{ii}]) shows a similar pattern, whereby the higher ionization lines 
like [\ion{Fe}{vii}] show a different morphology. In the low ionization lines 
the bright component A is the dominant source in the $l-v$--map. Comparing 
this with the velocity field in the [\ion{Fe}{vii}] line shows that this 
component is not visible. Here only a central component is appearing which is 
most likely identified with the nucleus of the Seyfert galaxy. Hence, there 
is some evidence that the true nucleus has a much lower apparent luminosity 
than the NLR component A, indicating a hidden nucleus. This picture is 
supported by HST imaging studies. It has clearly shown, that nuclear dust 
lanes which are also visible in NGC\,1386 are a common feature with typical 
scale heights of 20--50\,pc (Capetti et al. \cite{Capetti2}) which often 
leads to obscuration of the central region of an AGN. Furthermore this is 
indeed evidenced by X--ray measurements, which show that the active nucleus 
in NGC\,1386 is covered by a Compton--thick (i.e. $\rm{N_H} > 10^{24}\,\rm{
cm^{-2}}$) screen (Maiolino et al. \cite{MSB}).

\begin{figure}[]
\psfig{figure=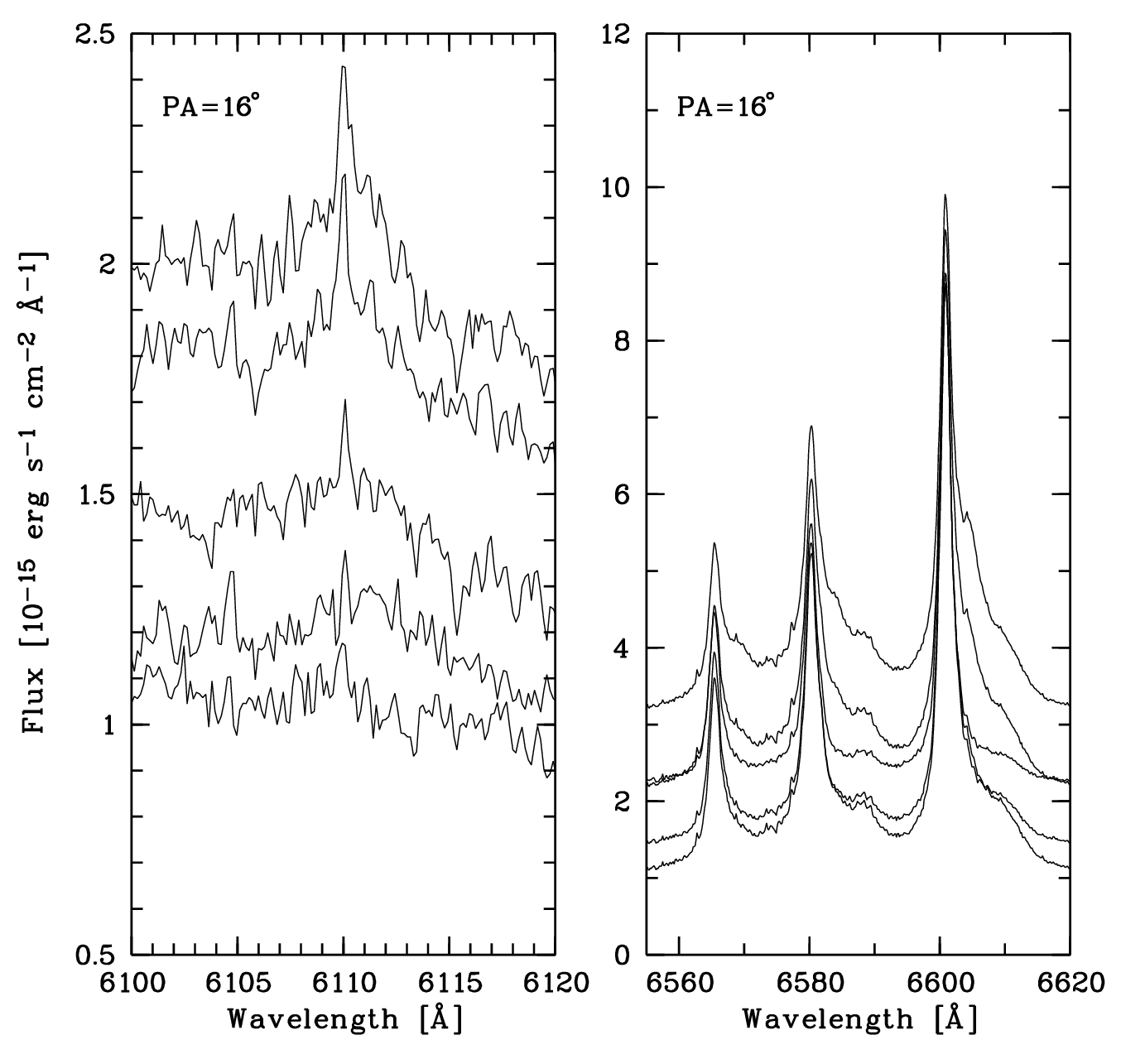,width=8.8cm,clip=t}
\caption[]{In the left panel the 5 spatial scans (each separated by 0\farcs65) 
of the [\ion{Fe}{vii}]\,$\lambda$\,6087\,{\AA} line are plotted and in 
comparison in the right panel the H$\alpha$+[\ion{N}{ii}] complex is 
shown for the same spatial scans. The observed wavelength is plotted 
versus the flux in units of $10^{-15}\,\mathrm{erg\,s^{-1}\,cm^{-2}}$\,{\AA}
$^{-1}$. To better illustrate the differences in the H$\alpha$+[\ion{N}{ii}] 
line scans, we have offset each individual scan by $5\times10^{-16}
\,\mathrm{erg\,s^{-1}\,cm^{-2}}$\,{\AA}$^{-1}$, and in the [\ion{Fe}{vii}]\,
$\lambda$\,6087\,{\AA} line by $2\times10^{-16}\,\mathrm{erg\,s^{-1}\,
cm^{-2}}$\,{\AA}$^{-1}$}
\label{F4}
\end{figure}

In Fig.~\ref{F4} we show the spatial variation of profiles of the 
[\ion{N}{ii}], H$\alpha$ lines together with the 
[\ion{Fe}{vii}]\,$\lambda$\,6087\,{\AA} line. 

\vspace{0.5cm}
\vspace{0.5cm}

\subsection{Identification of individual NLR components}

\begin{figure*}[t]
\psfig{figure=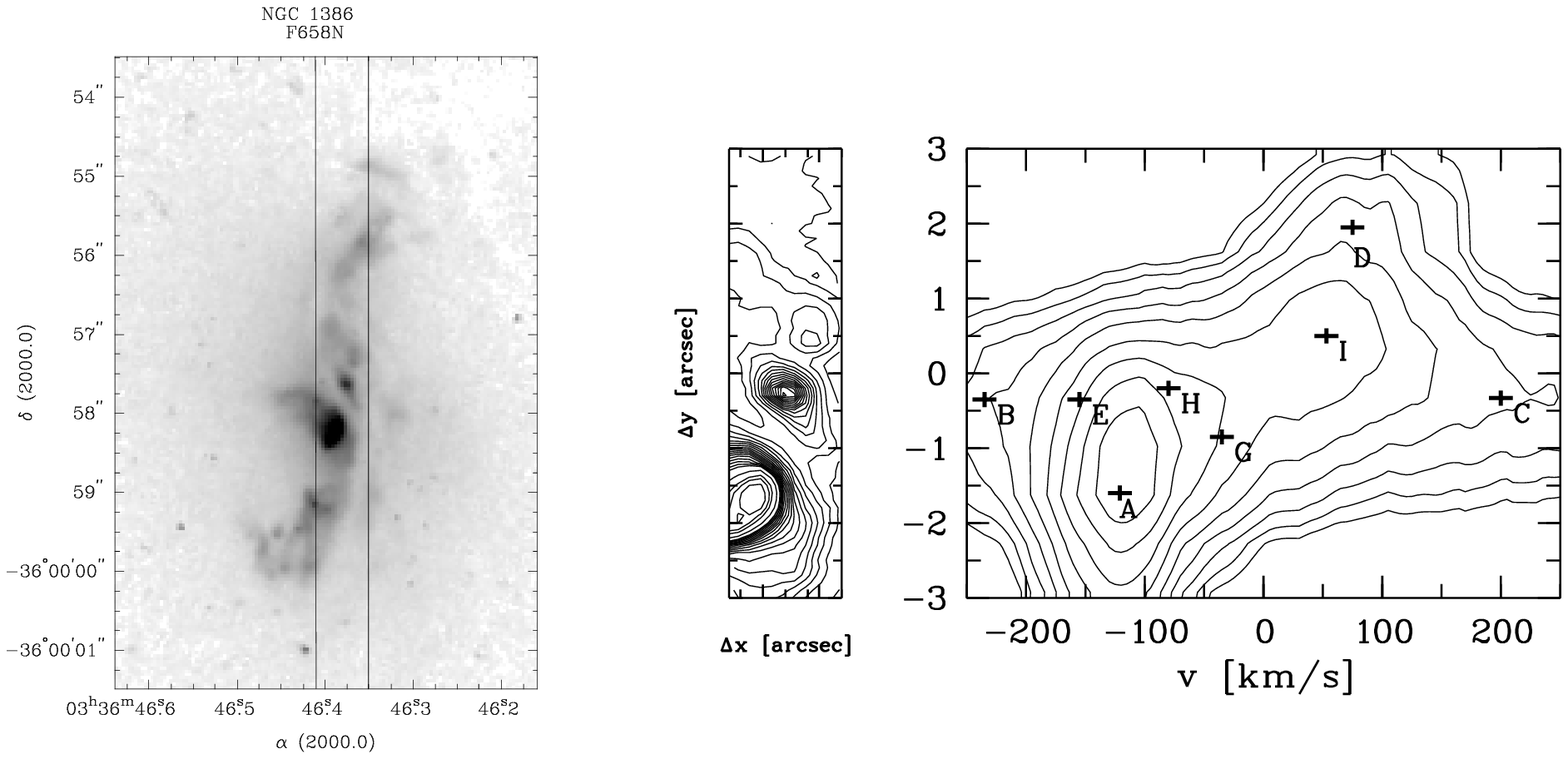,width=17cm,clip=t}
\caption[]{Comparison of our CASPEC $l-v$ map with the HST H$\alpha$ image. 
The left panel (5\,a) shows the central region of the WFPC\,2 (F658N) 
image of NGC\,1386 (rotated clockwise by 16\degr) with the CASPEC slit 
superimposed. The second sub--panel (5\,b) shows the central region of the 
F658N image in contours which was covered by the central 6\arcsec of the slit 
of our echelle spectrum. The right panel (5\,c) is the $l-v$ map of the 
[\ion{N}{ii}]\,$\lambda$6583\,{\AA} line (shown in logarithmic scale) where 
the velocity relative to the systemic velocity is plotted versus the radius 
from the center in arcseconds. The positions of the individual NLR components 
(as derived from the Gaussian decomposition) are marked by crosses and are 
accompanied by the respective letters. Due to the large separation in 
velocity space we have omitted the component F in the map. The accurracies 
of the velocities ($10\,{\rm{km\,s^{-1}}}$), and of the positions (0\farcs1) 
are indicated by the size of the crosses.} 
\label{F5}
\end{figure*}

Next we made a detailed comparison of our reduced 2-D echelle spectrogram 
with WFPC\,2 archive images taken with the HST. In Fig.~\ref{F5}\,a we show 
a superposition of the inner part of NGC\,1386 of the F658N filter HST image 
with the CASPEC slit. In Fig.~\ref{F5}\,b we show the region (in contours) 
which is covered by the CASPEC slit. In the right--hand panel of 
Fig.~\ref{F5} the direct comparison of our CASPEC spectrum of the 
[\ion{N}{ii}]\,$\lambda$\,6583\,{\AA} line as a $l-v$--map is shown. The 
contours are given in logarithmic scale. The outer two spectral scans of our 
echelle spectra are severely affected as a result of the flatfielding process 
as described in Sect. 2. We limit our study to the central $6\arcsec$, where 
only the outermost regions are to some point influenced by the flatfield 
uncertainties on a small scale. 

Now we can identify the individual NLR components seen in our echelle 
spectrogram with the components in the WFPC\,2 image. Due to the high 
dispersion individual components of line emission are separated in velocity 
space and can be located spatially with high precision (as discussed in 
Dietrich \& Wagner \cite{DW}). Therefore, we can cross--check the components 
(or emission knots) in the two `images'. This investigation of the morphology 
can also be regarded as an additional method to the direct imaging studies in 
the respective narrow--bands (e.g.,\,[\ion{O}{iii}], H$\alpha$). 

In Fig.~\ref{F5}\,c we have plotted the velocities of the individual NLR 
components that have been identified in our [\ion{N}{ii}]\,
$\lambda$6583\,{\AA} $l-v$--map (after Gaussian decomposition) as a function 
of the distance to the nucleus in arcseconds. The error bars arise from the 
uncertainties in the determination of the $l-$ and $v-$ coordinates in our 
$l-v$ map. 

The velocity field of NGC\,1386 is not as much disturbed as the velocity field 
in NGC\,1068 (Wagner \& Dietrich \cite{WD}) which shows velocity differences 
up to $1000\,\mathrm{km\,s^{-1}}$. Weaver et al. (\cite{WWB}) tried to 
determine the position of the nucleus in their spectroscopic study. Since 
their spectra were taken at medium resolution, the position of the nucleus of 
NGC\,1386 could not be identified unambiguously. Most likely a definitive 
answer would be given by polarimetric studies, such as it was done in the 
case of NGC\,1068 (Capetti et al. \cite{Capetti1}, \cite{Capetti3}), or by 
astrometry of the radio nucleus.  

The brightest component in our echelle spectrum coincides with the brightest
component in the WFPC\,2 image of NGC\,1386. This bright component A is 
offset by $\approx$95\,pc with respect to the nucleus. A comparison of 
the HST WFPC\,2 image (F658N) with our $l-v$ map of [NII]\,$\lambda$6583 
suggests a correspondence of the components A and G with the bright dominating 
component at $\sim-1\farcs5$ in the HST image of NGC\,1386. These two 
components have a similar velocity dispersion of FWHM$\approx$100\,km\,s$^{
-1}$ and are blueshifted with respect to the nucleus. Our components D and I 
are associated with the emission knots at $\approx2\farcs0$ and $0\farcs5$ in 
the HST image, respectively. The FWHM of D and I is comparable to the FWHM of 
A,G, and H, however the components D and I are redshifted with respect to the 
nucleus. The components B,C,E,H might be connected with the central 
bright emission complex at $\approx-0\farcs3$ in the WFPC\,2 image. In 
contrast to the other components they are nearly located along the line of 
sight and show a velocity dispersion of the order of 
FWHM\,$\approx$\,350\,km\,s$^{-1}$. The velocity split of the individual 
components is of the order of 400\,km\,s$^{-1}$ and taking the weak and quite 
uncertain component F into account it rises to 
$\Delta v\simeq$\,800\,km\,s$^{-1}$. Due to the large separation in 
velocity space from the other components we have not plotted component F into 
the $l-v$ map.

\vspace{0.5cm}
\vspace{0.5cm}


\section{Discussion}

One of the motivations for the investigation of the velocity field of the NLR 
in NGC\,1386 is to examine whether there are correlations between radio--
structures and optical emission. In the theoretical picture the deviations 
from a pure rotational component of the velocity are manifested in infall or 
outflow motions. In the case of an outflow motion, this often leads to the 
detection of corresponding radio structures in the highly resolved VLA maps. 
Those structures are mostly identified with a jet, propagating through the 
ambient medium. The interaction of the outflowing radio plasma with the NLR 
gas will produce shocks and this shock gas can be described as an expanding 
cooling cocoon around the radio jet (Taylor et al. \cite{TDA}; Axon et al. 
\cite{Axon}). Evidence for this scenario is found by Axon et al. (\cite{Axon}) 
who mapped spectroscopically the inner few arcseconds of NGC\,1068 at high 
resolution. They detected a velocity split of the order of 1000 km\,s$^{-1}$ of
individual components which was also found by Dietrich \& Wagner (\cite{DW}).
The observed velocity difference in NGC\,1386 can be explained within the 
framework of the models suggested by Taylor et al. (\cite{TDA}) and Axon 
et al. (\cite{Axon}).  

The slit position of NGC\,1386 in our investigation was chosen because the 
study by Mauder et al. (\cite{Mauder}) showed that the NLR has an elongated 
structure, with an asymmetrical biconical shape. A distinct substructure is 
visible in their speckle images and a few components could be resolved in the 
central part (100\,pc). From their speckle imaging an opening angle could be 
derived and the extended H$\alpha$--emission is mostly visible at a position 
angle of $\sim 20\degr$. Our identifications of the NLR components are in 
agreement with the observed NLR morphology by Mauder et al. (\cite{Mauder}). 

Our simple approach to reconstruct the H$\alpha$, [NII] emission line
complex using elliptical Gaussian profiles provide some evidence for two
components of the NLR velocity field of NGC1386. Our components can be
associated with the bright emission line complexes of an HST narrow-band
image (F658N). The components which might be dominated by the general
galactic rotation can be characterized by FWHM\,$\approx$\,100\,km\,s$^{-1}$
while the components with FWHM\,$\approx$\,350\,km\,s$^{-1}$ might be in radial
motion. This result is in good agreement with Weaver et al. (\cite{WWB}) who
also provide evidence for a two component velocity field of NGC\,1386. 

Even the velocity difference of the components with larger FWHM is of the
order of 400 km\,s$^{-1}$ only, there might be some interaction of the NLR
gas with the outflowing radio plasma. The study by Ulvestad \& Wilson 
(\cite{UW}) revealed a slightly elongated structure in their VLA map 
($\lambda$=6\,cm) of NGC\,1386 at the p.a. $225\degr$. So the radio--emission 
points anti-parallel to the optical emission. However, a more recent 
investigation by Nagar et al. (\cite{NWMG}) revealed a 300\,pc extended 
feature in p.a. $\sim170\degr$ in their VLA $\lambda3.6$\,cm map. 

A further interesting result of our analysis is the detection of the bright
component (cloud A) which is displaced by $-1\farcs6$ from the nucleus. In 
the study by Weaver et al. (\cite{WWB}) they reported a displacement of the 
optical nucleus by $\sim1\farcs5$ from their kinematical studies (kinematical 
center), which is most likely coincident with cloud A in our investigation. 
The true nucleus of NGC\,1386 has a much less apparent luminosity than cloud 
A, and hence is most likely highly absorbed. This is also evidenced by 
X--ray observations (Maiolino et al. \cite{MSB}). We conclude that component 
A is excited by photoionization, as is evident from the measured line 
ratios.  

The fact that there is a cone--like geometry of the NLR on small scales,
hints that absorption plays a role on those scales and that effect implies 
that the ionizing radiation can not escape isotropically from the 
circumnuclear regions. The observation of cone--like geometries in the 
central regions of Seyfert galaxies is a common feature. Evans et al. 
(\cite{Evans}) have shown that the NLR in NGC\,1068 is cone--like and 
comprises of several components. The kinematics of the strong components of 
this cloud ensemble have been studied by Dietrich \& Wagner (\cite{DW}).
At least $\sim$16 components are visible in their $l-v$--map obtained for the 
[\ion{O}{iii}]\,$\lambda$\,5007\,{\AA}--line down to the scale of $0\farcs2$ 
that reside in the central $6\arcsec$ region. From their study it was inferred 
that the velocity field of NGC\,1068 is driven by shocks due to the interaction
of the radio jet with the ambient medium. 


\section{Summary and Conclusions}

We have studied the kinematics and the morphology of the NLR in the Seyfert
galaxy NGC\,1386 with high spatial resolution with CASPEC and archival HST
images. The following results have been obtained.

\begin{itemize}
\item the NLR of NGC\,1386 is composed of 9 individual components with 5 
      components being redshifted and 4 being blueshifted with respect to 
      the systemic velocity.
\item the brightest component (cloud A) is blueshifted by $120\pm10\,
      \mathrm{km\,s^{-1}}$ and shows an offset of $1\farcs6$ to the south.
\item this cloud A is identified with the brightest region in the HST 
image, and is excited by photoionization
\item the true nucleus NGC\,1386 has a much lower apparent luminosity, and 
is possibly hidden by nuclear dust lanes.  
\item there is evidence for a two component velocity field in NGC\,1386
\end{itemize}

The results show that with this technique of sub--arcsecond mapping, which  
makes use of highly resolved 2-D spectra, it is possible to identify 
individual NLR components in the nearest Seyfert galaxies. In this 
respect it is a powerful tool to describe the kinematics and morphology of 
the NLR.

\begin{acknowledgements}We would like to thank Dr.~Andreas J\"uttner who 
carried out the spectroscopic observations. We thank Dr.~Hartmut Schulz for 
valuable comments on the manuscript. The Space Telescope Science Institute 
(STScI) is acknowledged for making the NGC\,1386 images available to us 
through their HST--data archive. This research was supported by the DFG 
(Deutsche Forschungsgemeinschaft) through SFB\,328. In addition, JR 
acknowledges financial support from the DLR (Deutsches Zentrum f\"ur Luft-- 
und Raumfahrt) through grant 50 0R 9707. MD is supported through NASA grant 
NAG\,5-3234.
\end{acknowledgements}


\end{document}